\documentclass[epj]{svjour}
\usepackage{graphicx}

\usepackage{rotating}
\usepackage[T1]{fontenc}
\usepackage[latin1]{inputenc}

\def\lapprox{{\raise0.5ex\hbox{$<$}\hskip-0.80em\lower0.5ex\hbox{$\sim$}

}}

\def\gapprox{{\raise0.5ex\hbox{$>$}\hskip-0.80em\lower0.5ex\hbox{$\sim$}

}}

\usepackage{graphics}
\begin{document}

\title{Single-Pion Production in $pp$ Collisions at 0.95 GeV/c (II)}  

\author{
S.~Abd El-Samad\inst{8}\and
R.~Bilger\inst{6}\and%,
K.-Th.~Brinkmann\inst{2}\and
H.~Clement\inst{6} \and
M.~Dietrich\inst{6}\and
E.~Doroshkevich\inst{6}\and
S. Dshemuchadse\inst{5,2}\and
K.~Ehrhardt\inst{6}\and
A.~Erhardt\inst{6}\and
W.~Eyrich\inst{3}\and
A.~Filippi\inst{7}\and
H.~Freiesleben\inst{2}\and
M.~Fritsch\inst{3,1}\and
R.~Geyer\inst{4}\and
A.~Gillitzer\inst{4}\and
J.~Hauffe\inst{3}\and
D.~Hesselbarth\inst{4}\and
R.~Jaekel\inst{2}\and
B.~Jakob\inst{2}\and
L.~Karsch\inst{2}\and
K.~Kilian\inst{4}\and
J. Kress\inst{6}\and
E.~Kuhlmann\inst{2}\and
S.~Marcello\inst{7}\and
S.~Marwinski\inst{4}\and
R.~Meier\inst{6}\and
K. M\"oller\inst{5}\and
H.P. Morsch\inst{4}\and
L.~Naumann\inst{5}\and
J.~Ritman\inst{4}\and
E.~Roderburg\inst{4}\and
P. Sch\"onmeier\inst{2,3}\and
M. Schulte-Wissermann\inst{2}\and
W.~Schroeder\inst{3}\and
F. Stinzing\inst{3}\and
G.Y. Sun\inst{2}\and
J.~W\"achter\inst{3}\and
G.J.~Wagner\inst{6}\and
M.~Wagner\inst{3}\and
U.~Weidlich\inst{6}\and
A. Wilms\inst{1}\and
S.~Wirth\inst{3}\and
G.~Zhang\inst{6}\thanks{present address: Peking University}\and
P. Zupranski\inst{9}
}
%
%\offprints{H. Clement}          % Insert a name or remove this line
\mail{H. Clement \\email: clement@pit.physik.uni-tuebingen.de}
%\email{clement@pit.physik.uni-tuebingen.de}
%

\institute{
Ruhr-Universit\"at Bochum, Germany \and
Technische Universit\"at Dresden, Germany \and
Friedrich-Alexander-Universit\"at Erlangen-N\"urnberg, Germany \and
Forschungszentrum J\"ulich, Germany \and
Forschungszentrum Rossendorf, Germany \and
Physikalisches Institut der Universit\"at T\"ubingen, T\"ubingen, Germany \and
University of Torino and INFN, Sezione di Torino, Italy \and
Atomic Energy Authority NRC Cairo, Egypt \and
Soltan Institute for Nuclear Studies, Warsaw, Poland 
\\
(COSY-TOF Collaboration)}
\date{\today}
%
%\date{Received: \today / Revised version: date}
%\date{Received: July 10, 2006}
\abstract{
The single-pion production reactions $pp\rightarrow d\pi^+$, $pp\rightarrow
np\pi^+$  and $pp\rightarrow pp\pi^0$ were measured at a beam momentum of
0.95 GeV/c ($T_p \approx$ 400 MeV) using the short version of the COSY-TOF
spectrometer. The central calorimeter provided particle identification,
energy determination and neutron detection in addition to time-of-flight and
angle measurements from other detector parts. Thus all pion production
channels were recorded with 1-4 overconstraints. The main emphasis is put on
the presentation and discussion of the $np\pi^+$ channel, since the results on
the other channels have already been published previously.
The total and differential cross sections  obtained are compared to
theoretical calculations. In contrast to the $pp\pi^0$ channel we observe in
the $np\pi^+$ channel a strong influence of the $\Delta$ excitation.
% already at this energy close to threshold. 
In particular the pion angular distribution exhibits a $(3~cos^2\Theta + 1)$
dependence, typical for a pure {\it s}-channel
$\Delta$ excitation and identical to that observed in the
$d\pi^+$ channel. Since the latter is understood by a {\it s}-channel
resonance in the $^1D_2$ $pn$ partial wave, we discuss an analogous scenario
for the $pn\pi^+$ channel. 
}
\maketitle
\section{Introduction}
\label{intro}

Single-pion production in the collision between two nucleons is thought to be
the simplest inelastic process between two baryons, nevertheless its
understanding is still far from being satisfactory - both from the theoretical
and the experimental point of view. In a recent publication \cite{evd} -- in
the following denoted by (I) -- we have
presented the first kinematically complete measurement for the $pp \to pp\pi^0$
channel at a beam momentum of 0.95 GeV/c (corresponding to $T_p=$397
MeV). Although $\Delta$ production in a relative s-wave is prohibited in this
reaction channel, we have seen that the angular distributions indicate the
presence of significant $\Delta$ production in a relative p-wave already close
to threshold.

In this work we present our results for the $pp \to pn\pi^+$ channel at the
same incident energy. We will show that here, where $\Delta$ production in
relative s-wave is allowed, indeed this production process is
overwhelmingly dominant and characterizes the differential observables in this
channel. 

The reaction of interest here has been investigated in the near-threshold
region already 
previously by a series of measurements at JINR \cite{neg}, KEK
\cite{shim,shim1}, TRIUMF 
\cite{pley}, CELSIUS \cite{AB} and notably at IUCF \cite{dae,har,flam,WD},
since  there experiments 
have been carried out both with polarized beam and target. In these
measurements very close to threshold it was already noted \cite{dae} that at
beam energies of $T_p$ = 320 MeV, i.e. only 20 MeV above threshold, an onset
of the $\Delta$ excitation is seen in the pion angular distribution. The latter
changes from isotropic at energies below 300 MeV to anisotropic at 320 MeV
suggesting there a 30 - 40 $\%$ contribution of the $\Delta$ excitation.
In the TRIUMF
measurements, which were performed at beam energies of $T_p$ = 420 and 500
MeV, it was noticed that at these energies the pion angular distributions of
$pp \to d\pi^+$ and $pp \to pn\pi^+$ reactions are not only very similar, 
but also very close to the angular distribution expected from a pure $\Delta$
excitation. In terms of partial waves this means a predominance of the $^1D_2$
$pp$ partial wave at these energies as is in fact the outcome of the SAID
\cite{said} partial wave analysis for the $pp \to d\pi^+$ reaction. Hence
after presentation and discussion of our data we will confront them with
theoretical {\it t}-channel calculations (Fig.1 , top) as well as with {\it
  s}-channel calculations (Fig. 1, bottom). The latter account for the
striking $^1D_2$ partial wave dominance  and assume
for simplicity that all other partial waves give negligible contributions. In
both cases an excitation of the $\Delta$ resonance is assumed.

\begin{figure}
\begin{center}
\includegraphics[width=10pc]{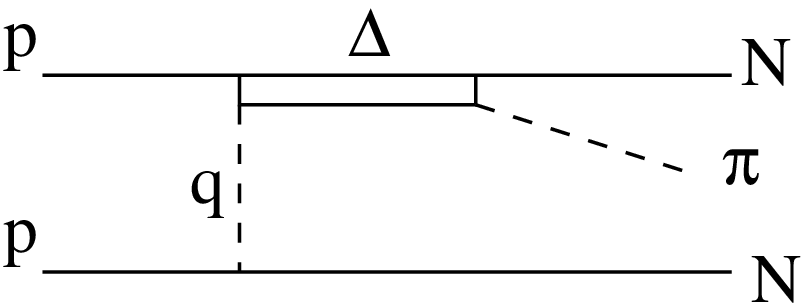}\\
\vspace{0.7 cm}
\includegraphics[width=10pc]{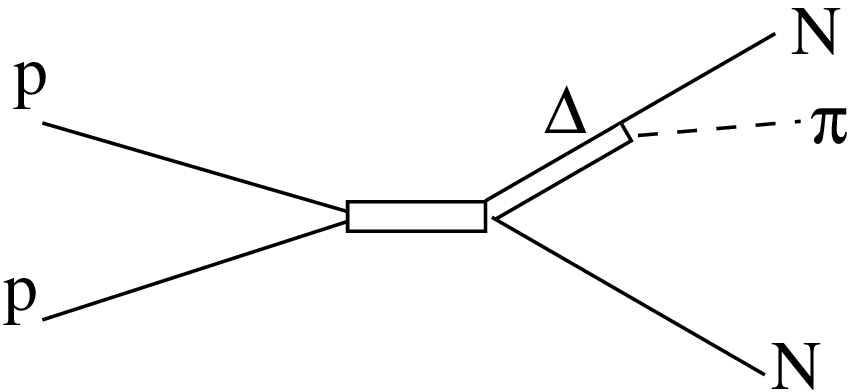}
\end{center}
\caption{Graphs used for the theoretical calculations. Top: t-channel
  approach (PWIA, direct part only), bottom: s-channel approach.   
 }
\end{figure}

\section{Experiment}
\label{sec:2}

\subsection{Detector setup}

Since the experimental setup has been discussed in detail already in (I), we
give here only a short account. 
The measurements have been carried out at the J\"ulich Cooler Synchrotron COSY
using the time-of-flight spectrometer TOF at one of its external beam
lines. At
the entrance of the detector system the beam - collimated to a diameter smaller
than 2 mm - hits the LH$_2$ target, which has a length of 4 mm, a diameter of
6 mm and 0.9 $\mu m$ thick hostaphan foils as entrance and exit windows. At a
distance of 22 mm downstream of the target the two layers of 
the start detector (each consisting of 1 mm thick scintillators cut into 12
wedge-shaped sectors) were placed followed by a two-plane fiber hodoscope
(96 x 96 fibers, 2 mm thick each ) at a distance of 165 mm from
target. Whereas the  start detector mainly supplies the start time signals for
the time-of-flight (TOF) 
measurements, the fiber hodoscope primarily provides a good angular resolution
for the detected particle tracks. In its central part the TOF-stop detector
system consists of the so-called Quirl, a 3-layer scintillator system  1081 mm 
downstream of the target --  and in its peripheral part of the so-called Ring,
also a 
3-layer scintillator system built in a design analogous to the Quirl, however,
with inner 
and outer radii of 560 and 1540 mm, respectively. Finally behind the Quirl a
calorimeter was installed for identification of charged
particles and of neutrons as well as for measuring the energy of charged
particles. The calorimeter 
consists of 84 hexagon-shaped scintillator blocks of length 450 mm, which
suffices to stop deuterons, protons and pions of energies up to 400, 300 and
160 MeV, respectively. The energy calibration of the calorimeter was performed
by detecting cosmic muons.

\subsection{Particle identification and event reconstruction}

In the experiment the trigger required two hits in the Quirl and/or Ring
associated with two hits in the start detector.
Tracks of charged particles are reconstructed from straight-line fits to the
hit detector elements. They are accepted as good tracks, if they
originate in the target and have a hit in each detector element the track
passes through. In this way the angular resolution is better than 1$^\circ$
both in azimuthal and in polar angles. If there is an isolated hit in the
calorimeter with no associated hits in the preceding detector elements, then
this hit qualifies as a neutron candidate (further criteria will be discussed
below). In this case the angular resolution
of the neutron track is given by the size of the hit calorimeter block,
i.e. by 7 - 8$^\circ$. By construction of the calorimeter a particle will hit
one or more calorimeter blocks. The number of blocks hit by a particular
particle is given by the track reconstruction. The total energy deposited by
this particle in the calorimeter is then just the (calibrated) sum
of energies deposited in all blocks belonging to the particular track.

In order to have maximum angular coverage by the detector elements and to
minimize the fraction of charged pions decaying in flight before reaching the
stop detectors, the short version of the TOF spectrometer
was used. In this way a total polar angle coverage of 3$^\circ \leq
\Theta^{lab}\leq$ 49$^\circ$ was achieved with the central calorimeter
covering the region 3$^\circ \leq\Theta^{lab}\leq$ 28$^\circ$. For fast
particles the energy resolution of the calorimeter amounting to 4\% is
superior to that from time-of-flight measurements due to the short path
length. However, the time-of-flight resolution is still much better than the
$\Delta E$ 
resolution of the Quirl elements. Hence, for particle identification, instead
of plotting $\Delta E$ versus 
$E_{cal}$, the uncorrected particle energy deposited in the calorimeter, we
utilize the relation  $\Delta E \sim (z/\beta)^2$ with the particle charge
$z=1$ and plot $1/\beta^{2}$ versus $E_{cal}$, where the
particle velocity $\beta=v/c$ is derived from the time-of-flight measurement.

By identifying and reconstructing the two charged tracks of an event the exit
channels $d\pi^+$, $np\pi^+$ and $pp\pi^0$ can be 
separated. Kinematically the maximum  possible laboratory (lab) polar angles
are $\approx 9^\circ$ for deuterons and $\approx 32^\circ$ for protons (and
neutrons). Hence 86\% of the angular coverage for protons and neutrons from
single pion production are within 
the angular acceptance of the calorimeter. For charged pions the angular
coverage has been much lower with this setup, since kinematically they can
extend up to $\Theta^{lab}~=$ 180$^\circ$. Hence within the angular coverage
of Quirl and Ring the angular acceptance for $\pi^+$ has been $\approx$40$\%$
only. Nevertheless most of the phase space part necessary for a full coverage
of the physics in single pion production has been covered (see below) by these
measurements due to the circumstance that the center-of-mass (cm) angular
distributions have to be symmetric about 90$^\circ$ because of identical
collision partners in the incident channel.

\subsection{Selection of the $np\pi^+$ channel}

The $np\pi^+$ channel is selected by identifying proton and pion
in the calorimeter or only the proton in the calorimeter, when the second
charged track is in the Ring. In addition, the missing $p\pi$ mass $MM_{p\pi}$
has to meet the condition 900 MeV$/c^2$ $\leq MM_{p\pi} \leq$ 980 MeV$/c^2$
. Also to 
suppress background from the $d\pi^+$ channel - in particular when the
deuteron breaks up and appears as a proton in the calorimeter - the $p\pi^+$
 track is required to be non-coplanar.To this end we determine the variable
$\Delta\Phi$, which is defined as the projection of the opening angle
  between two tracks onto the plane normal to the beam vector. That way we
  have always $\Delta\Phi \leq 180^\circ$. The according histogram is
  displayed in Fig.2. We see that $p\pi^+$ events stemming from deuteron
  breakup give rise to a large peak near $180^\circ$. 
%Its width of about
%  $5^\circ$ is given by the detector resolution in $\Phi$. 
In order to get rid
  of these events we introduce the constraint $\Delta\Phi < 160^\circ$. From
  Fig. 2 we see that in prinicple a cut $\Delta\Phi < 170^\circ$ would already
  be sufficient to eliminate this background. However, in order to be on the
safe side and to have no perceptible tails from the $pp \to d\pi^+$ reaction in
  the final  data sample, we used the more rigorous constraint.  
 From Fig. 2 we see that this constraint does not cut
away significant pieces of physics information \footnote{we note that the data
  sample with the cut $\Delta\Phi < 170^\circ$ leads to results, which are
  practically indistinguishable from those shown here}. Any extrapolation
into the cut region, be it by model or by phase space, will introduce
uncertainties on the level of less than two percent.

Aside from this physics background due to deuteron breakup we do not find any
sizable background in the final data sample, as has been checked by control
measurements, where the target cell was empty.

\begin{figure}
\begin{center}
\includegraphics[width=20pc]{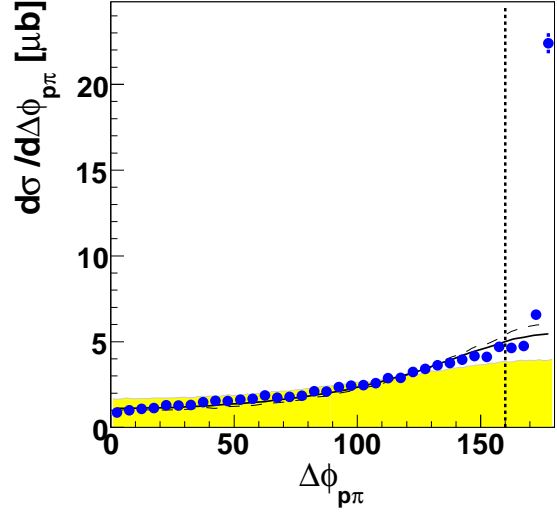}
\end{center}
\caption{Distribution of the planarity angle $\Delta\phi$ defined 
  in the center-of-mass system for the $pp\rightarrow pn\pi^+$
  reaction. Data of this work are shown by full circles and
  phase space by the
  shaded area. Solid and dashed lines denote s-channel and t-channel
  calculations, respectively, as discussed in the text. For ease of comparison
  the calculations have been normalized to the experimental total cross
  section. The constraint $\Delta\Phi < 160^\circ$ for the elimination of events
  stemming from deuteron breakup is indicated by the vertical dash-dotted line.
}
\end{figure}

Further on, the neutron 4-momentum is reconstructed from the 4-momenta of 
proton and pion and it is checked, whether a calorimeter block in the
corresponding ($\Theta$, $\Phi$) region recorded a hit without any additional
entries recorded in the preceding detector elements of the Quirl. If these
conditions are met, a neutron track is assumed. That way $\Theta_n$ and
$\Phi_n$ are determined by the location of this calorimeter block. By this
method we obtain a neutron detection efficiency of 36 $\%$. Thus having
only the neutron energy undetermined experimentally we end up with 3 kinematic
overconstraints for this channel. Corresponding kinematic fits were applied.

The luminosity of the experiment was determined from the analysis of $pp$
elastic scattering, see (I). 
All data have been efficiency corrected by MC
simulations of the detector setup by using the CERN GEANT3 \cite{geant}
detector simulation 
package, which accounts both for electromagnetic and hadronic interactions of
the ejectiles with the detector materials.

\subsection{Uncertainties}

The final data sample contains about 80000 good events, i.e. the
statistical 
errors  are on the percent level and hence of minor importance in comparison
with 
systematical uncertainties. One
source of systematic errors are beam alignment and quality. The requirement
that the beam has to pass a 2 mm collimator without significant halo assures
good alignment and quality of the beam. The alignment is also verified by the
fact that the angular distributions in the overall
center-of-mass system have to be symmetric about 90 $^\circ$, see next
section. In addition polarized beam measurements with the identical detector
setup used here also give no hint for noticeable misalignments \cite{AE}. 

A much
more severe source for systematic uncertainties concerns the acceptance and
efficiency corrections of the data. As discussed above the detector covers the
complete momentum range , however, not the complete angular range of the
reaction of interest. This means, that the acceptance correction has to rely on
extrapolations into unmeasured angular regions. From the Dalitz plots
displayed in Fig. 4 we see that -- according to all we know from this
reaction -- these unmeasured regions are the ones with the lowest cross
section. Hence systematic errors introduced by these extrapolations ought to
be of minor importance. To quantify this statement we compare the data
resulting from corrections with MC simulations, which are based either on pure
phase 
space or alternatively on a model description of the $pp \to n p \pi^+$
reaction. A model, which is trimmed to describe all essential features of the
data, is the appropriate tool for a reliable acceptance and efficiency
correction by a MC simulation, which passes the
ejectiles from the reaction of interest through the
virtual detector. It has the potential of providing a selfconsistent procedure
for these corrections.  Contrary to this the pure phase space description of
the reaction  is -- as we easily can see from the experimental results
displayed in Figs. 2 - 6 -- inadequate though convenient. It may serve,
however, as a very conservative estimate of the systematic uncertainties due to
the acceptance and efficiency corrections. As examples we display in Fig. 3
our data for the invariant $pn$ mass $M_{pn}$ and for the pion angular
distribution in the center-of-mass system $\sigma(\Theta_\pi^{CM})$ evaluated
(properly) by use of a model, which fits the final data (solid circles),
and alternatively by use of pure phase space (open circles), respectively. The
use of the latter for the extrapolation into unmeasured regions brings the
data, of course, somewhat 
closer to the phase space predictions. However, despite the fact that phase
space and model predictions are vastly different, the effect of the inadequate
phase space correction on the data is still very moderate. A realistic upper
limit for systematic uncertainties due to acceptance and efficiency correction
will be much smaller than the differences between open and solid circles in
Fig. 3. As a realistic estimate for this kind of systematic uncertainties we
conclude that they are within the size of the symbols, which are used in
Figs. 2 - 6 for displaying our experimental results. We note that the use
of either t-channel or s-channel calculations in the MC simulations does not
lead to any noticeable differences in the acceptance and efficiency corrected
data.  

\begin{figure}
\begin{center}
\includegraphics[width=10pc]{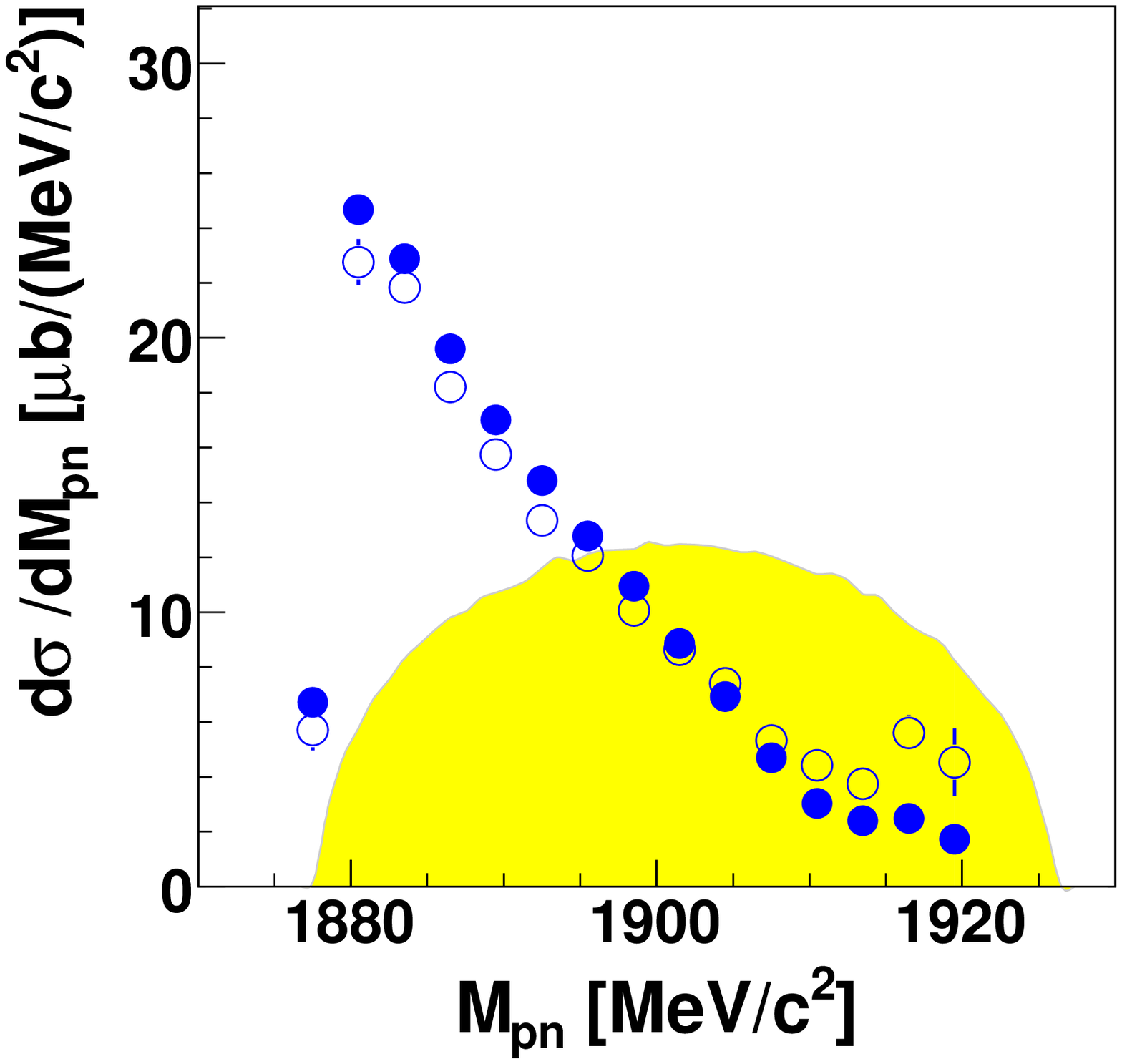}
\includegraphics[width=10pc]{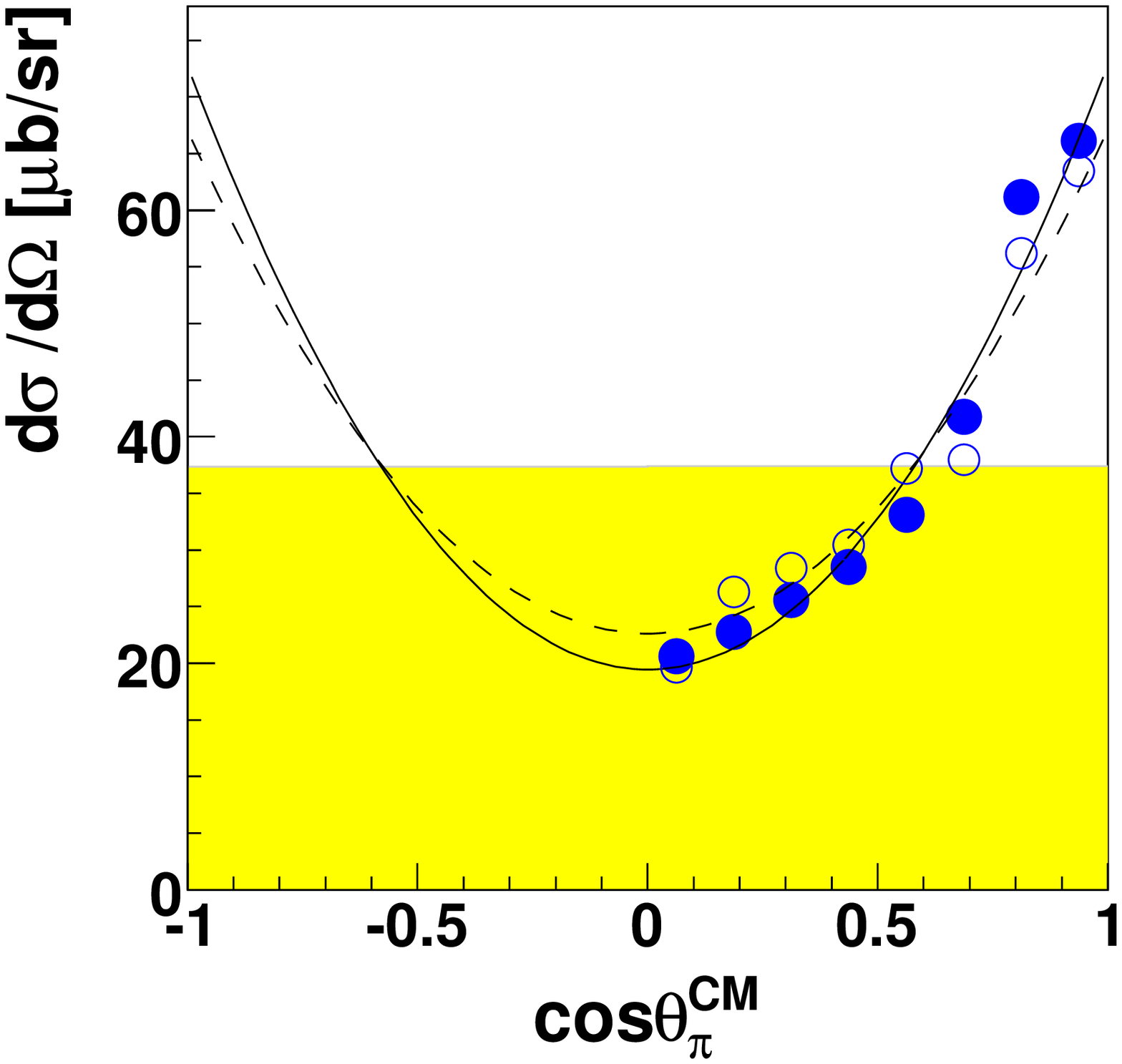}
\end{center}
\caption{Differential cross sections in dependence on the invariant mass
  $M_{pn}$ (left) and the pion scattering angle $\Theta_\pi^{CM}$ (right) in the
  center-of-mass system  for the $pp\rightarrow pn\pi^+$
  reaction. The full circles show our data properly corrected by a
  self-consistent MC simulation, i.e. using a reaction model, which is in
  agreement with the final data. The open circles derive from MC simulations
  using just pure phase space. The phase space distributions for the
  differential spectra are shown by the
  shaded area. Solid and dashed lines in the right figure show Legendre fits
  to the filled and open circles yielding $a_2$ = 0.96(2) and =0.79(2),
  respectively.  
}   
\end{figure}

%\textbf{3. Results}
\section{Results}
\label{sec:3}

Due to the identity of the collision
partners in the entrance channel the angular distributions in the overall
center-of-mass system have to be symmetric about 90 $^\circ$, i.e. the
full information about the reaction channels is contained already in the
interval $0^\circ\leq \Theta^{cm}\leq 90^\circ$. Deviations from this
symmetry in the data indicate systematic uncertainties in the
measurements. Hence we plot  - where appropriate - the full angular range, in
order to show the absence of major systematic errors in our
measurement.

The total cross section of 0.47(2) mb for the $pp\rightarrow pn\pi^+$ reaction
at $T_p = 400 MeV$ has already been given in (I). It is roughly a factor of
two smaller than that for the $pp \to d\pi^+$ channel and five times larger
than that for the $pp \to pp\pi^0$ channel, see Table 1 in (I). For the
comparison with data at neighbouring energies \cite{bys,shim,pley,har} see
discussion in the next section.

Differential distributions are shown in Figs. 3 - 6. Since a 3-body
system in the final state has five independent variables, we choose to present
the three invariant mass spectra $M_{pn}$, $M_{p\pi^+}$ and $M_{n\pi^+}$ as
well as the proton and pion angular distributions in the overall
center-of-mass system (cms). Because of the limited acceptance for pions only
the angular range $0^\circ < \Theta_{\pi}^{cm}\leq 90^\circ$ is
covered. However, as 
pointed out above, due to the required 90$^\circ$ symmetry of the angular
distributions the full information is contained in the data.

\begin{figure}
\begin{center}
\includegraphics[width=10pc]{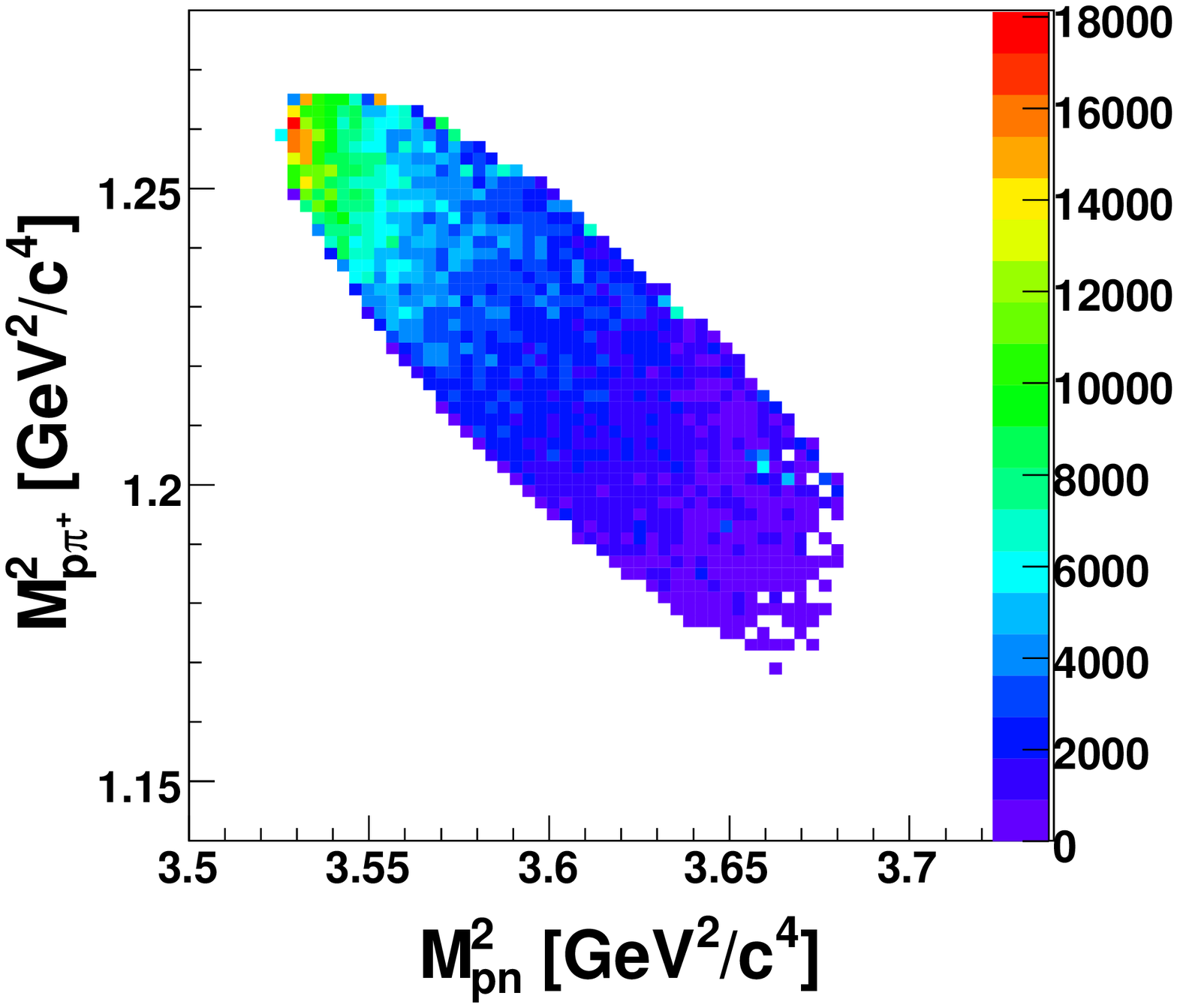}
\includegraphics[width=10pc]{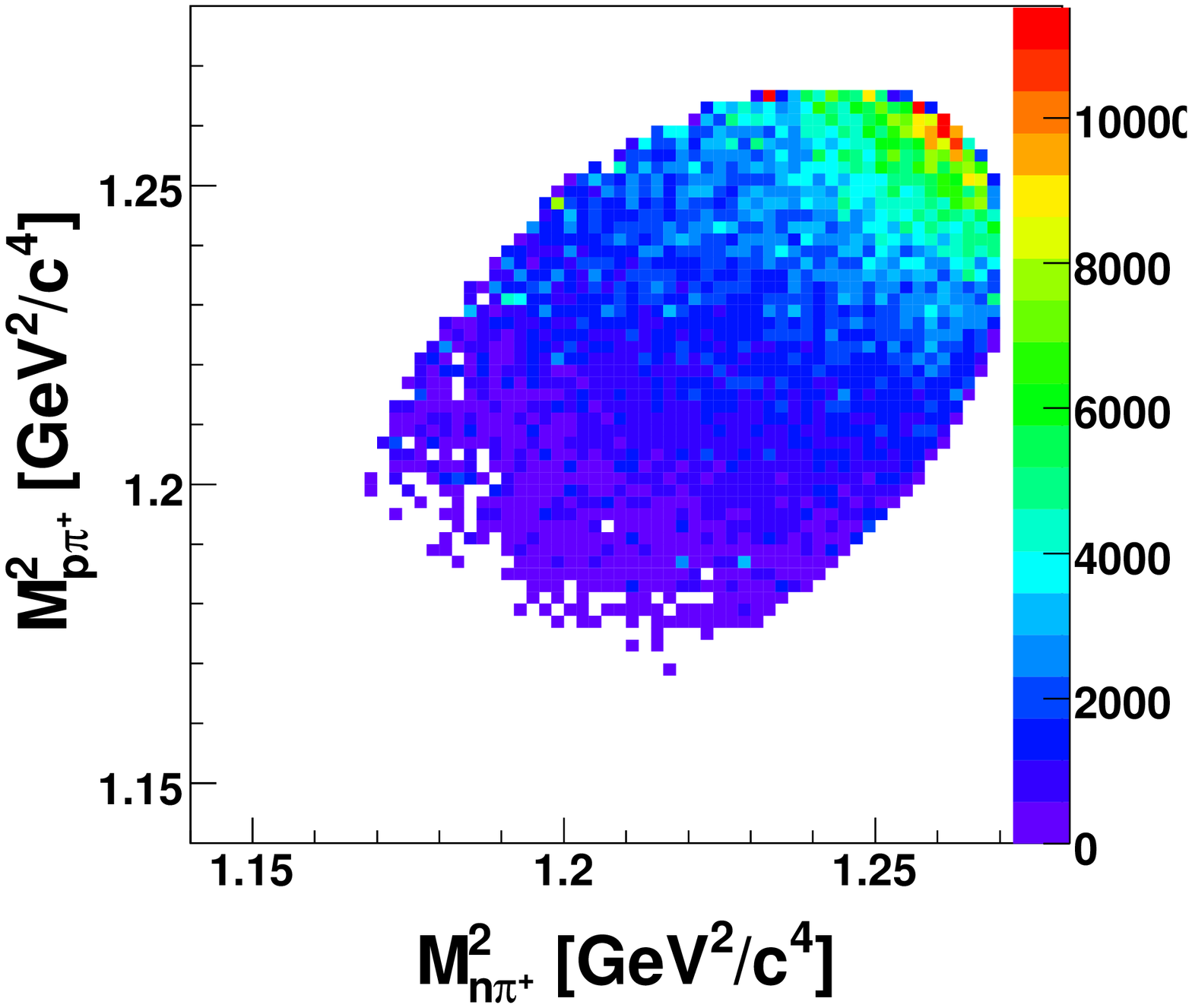}
\includegraphics[width=10pc]{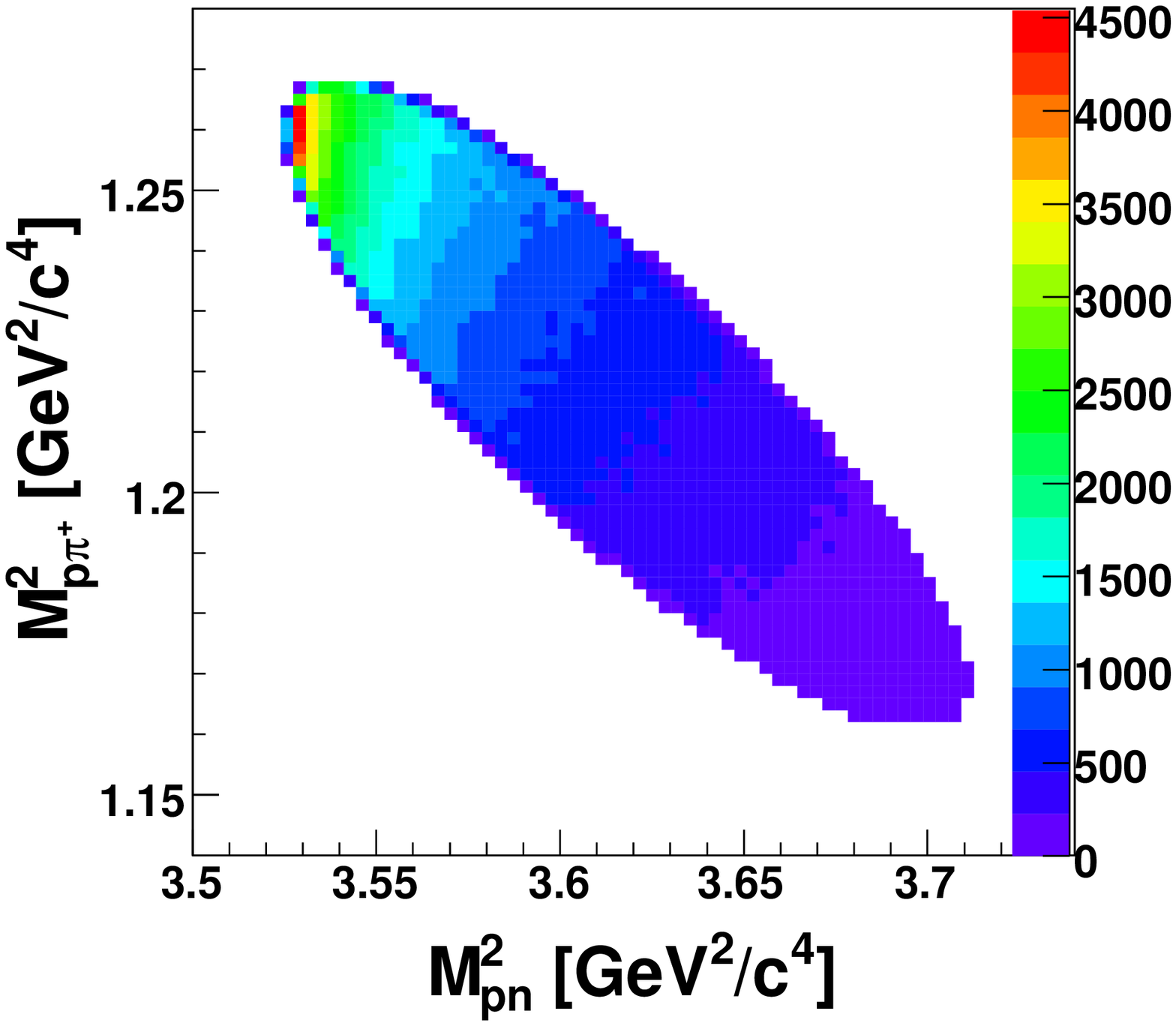}
\includegraphics[width=10pc]{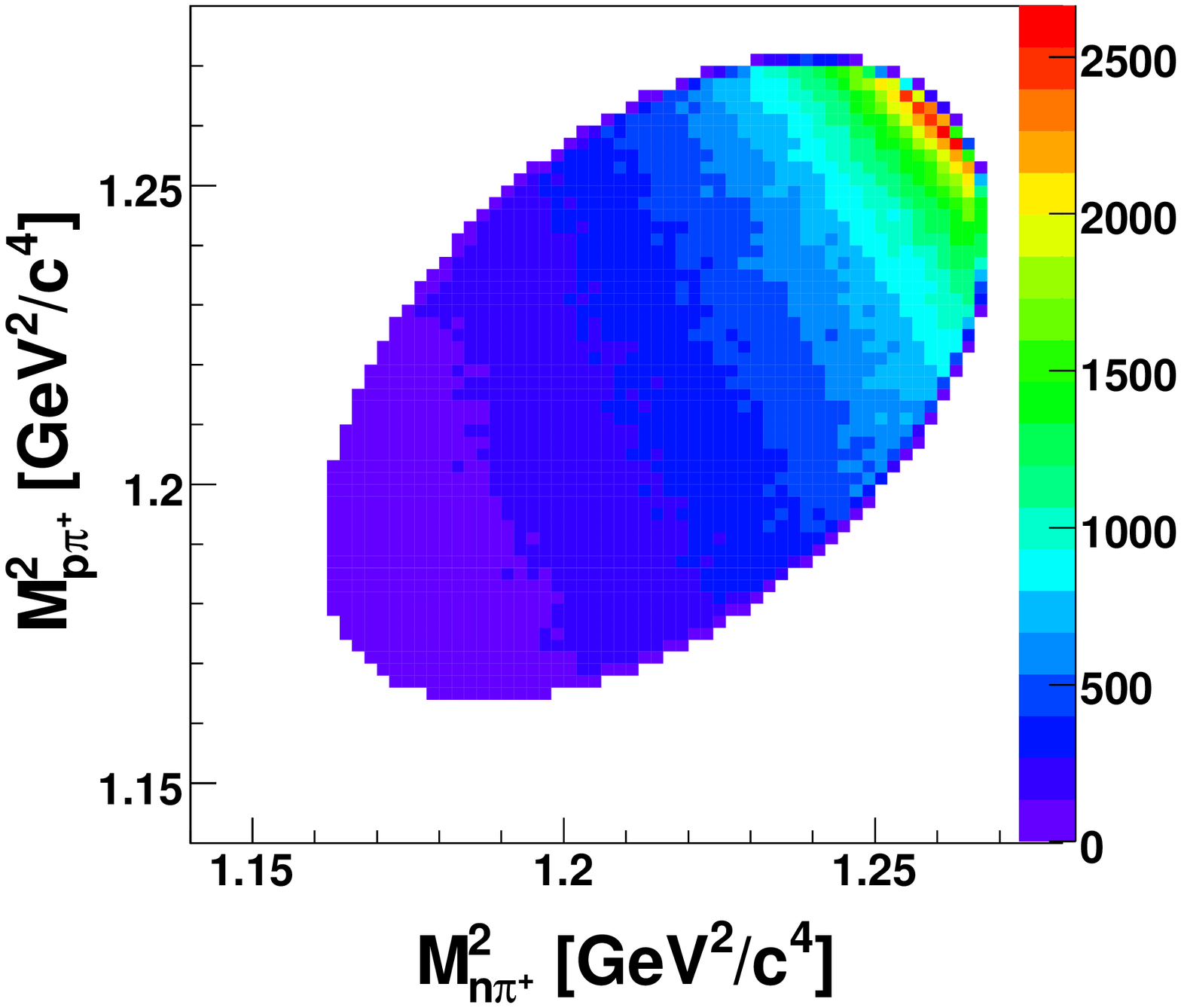}
\end{center}
\caption{Dalitz plots for the invariant mass combinations  $M_{p\pi^+}^2$ 
  versus $M_{pn}^2$ (left) and $M_{p\pi^+}^2$ versus $M_{n\pi^+}^2$ (right) as
  obtained for the $pp\rightarrow pn\pi^+$ reaction: data are shown on top and
  the MC s-channel calculation (see text) at the bottom. Note that the plots
  for the data are efficiency but not acceptance corrected. The tiny deviations
  from the elliptic circumference at the upper corners are due to the excluded
  beam-hole region.}
\end{figure}

\begin{figure}
\begin{center}
\includegraphics[width=10pc]{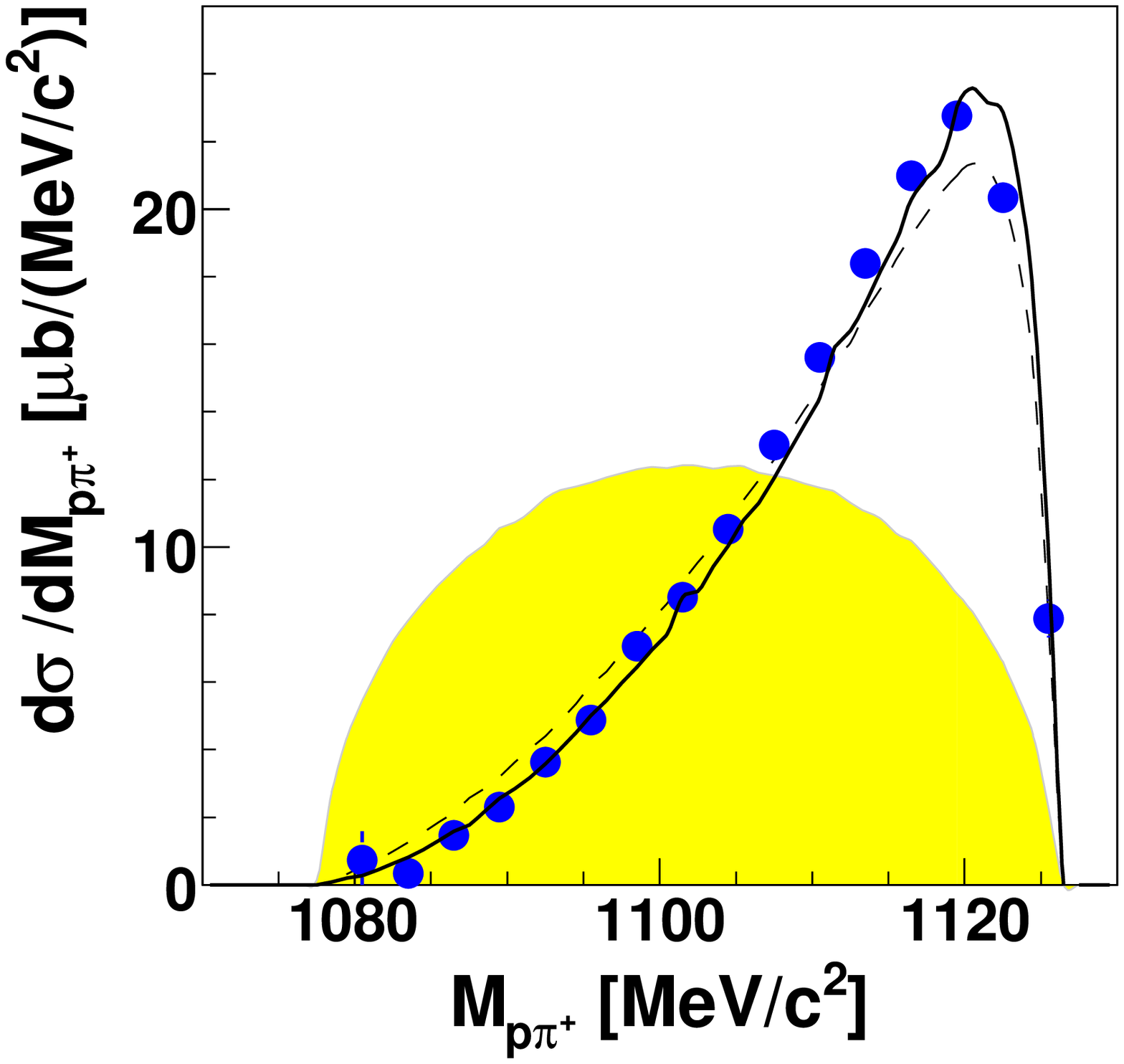}
\includegraphics[width=10pc]{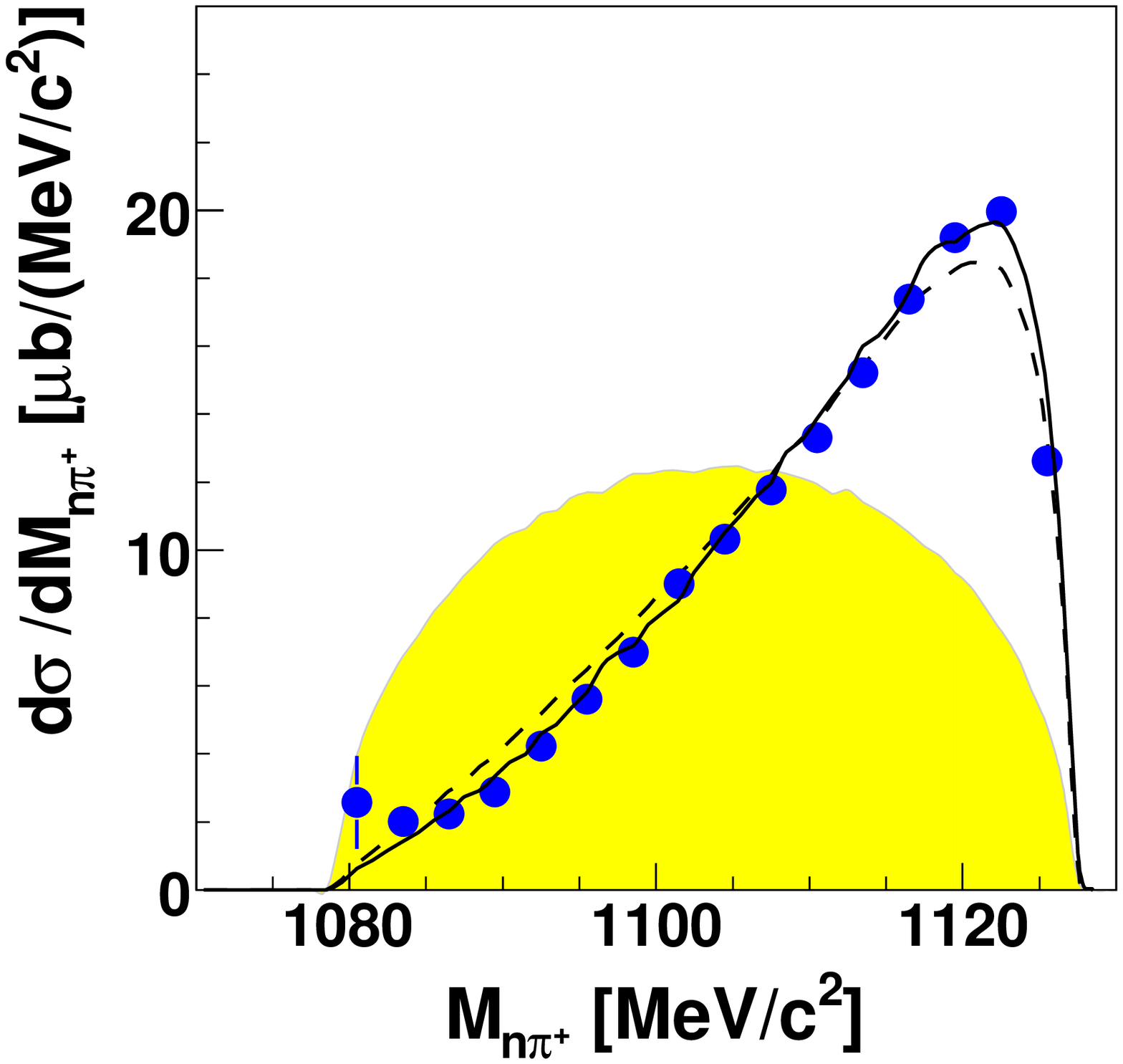}
\includegraphics[width=10pc]{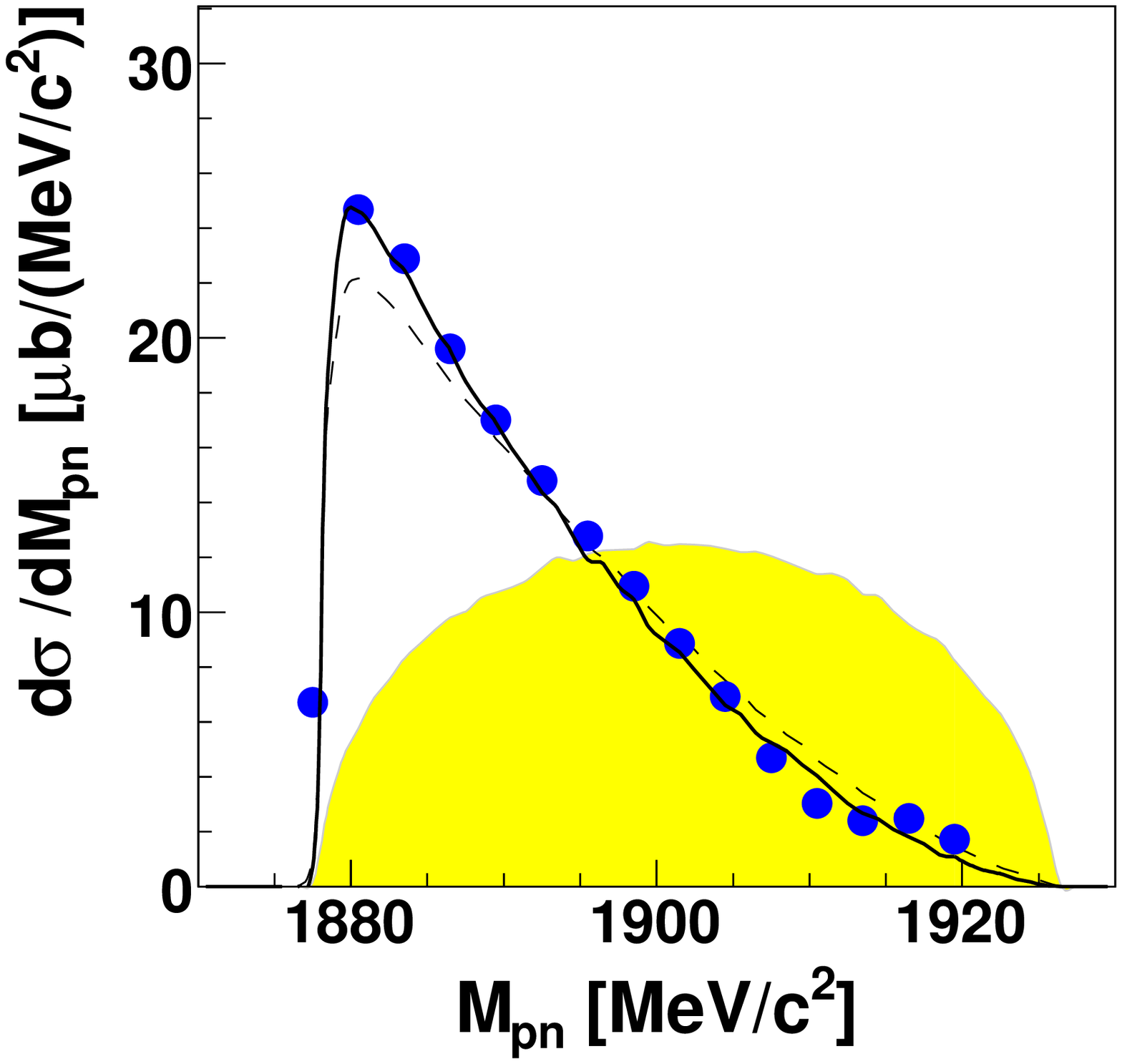}
\end{center}
\caption{Differential cross sections in dependence on invariant masses
  $M_{pn}$, $M_{p\pi^+}$ and $M_{n\pi^+}$ for the $pp\rightarrow pn\pi^+$
  reaction. Data of this work are shown by full circles and phase space by the
  shaded area. Solid and dashed lines denote s-channel and t-channel
  calculations, respectively, as discussed in the text. For ease of comparison
  the calculations have been normalized to the experimental total cross
  section.
}   
\end{figure}

\begin{figure}
\begin{center}
\includegraphics[width=10pc]{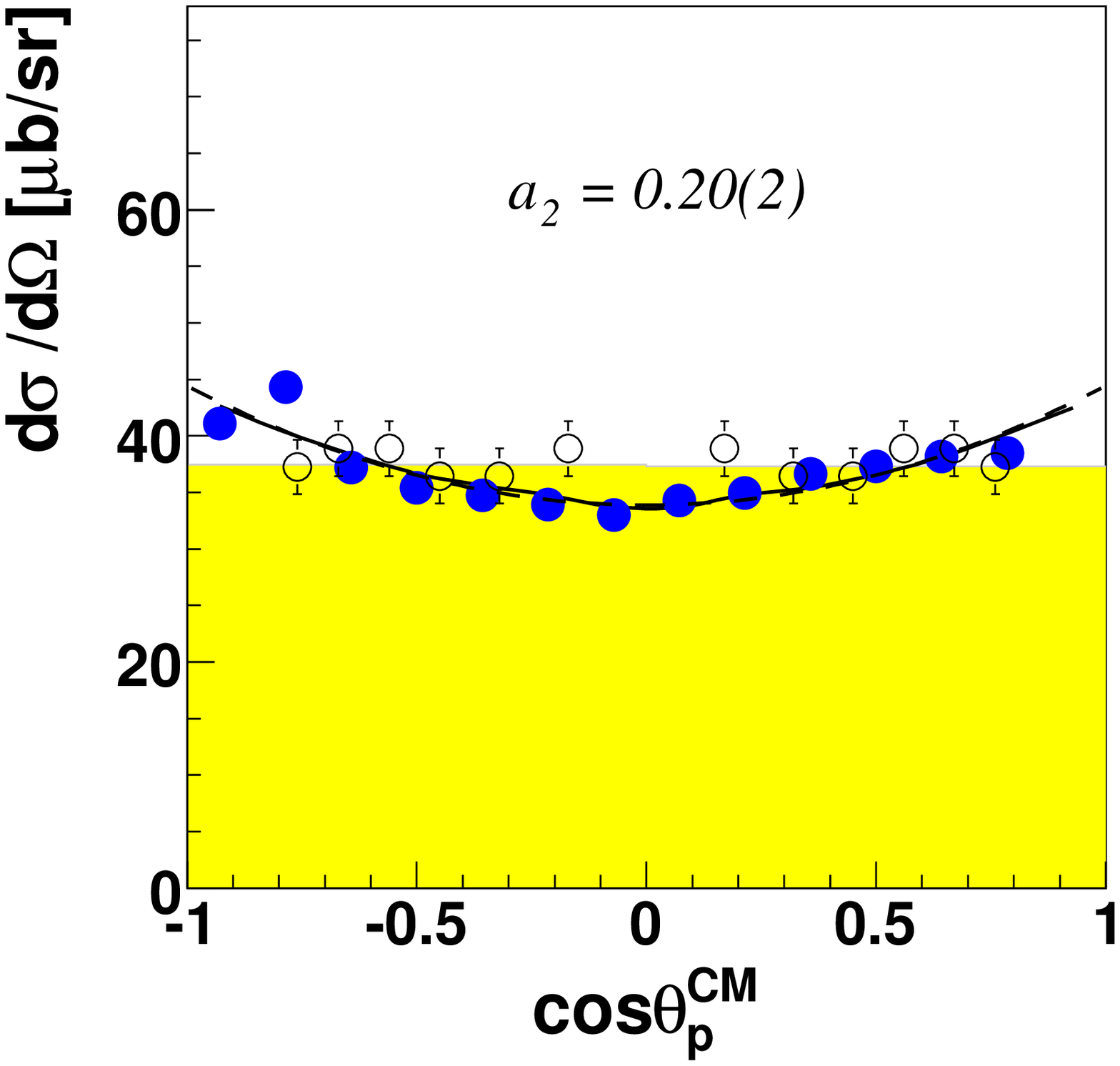}
\includegraphics[width=10pc]{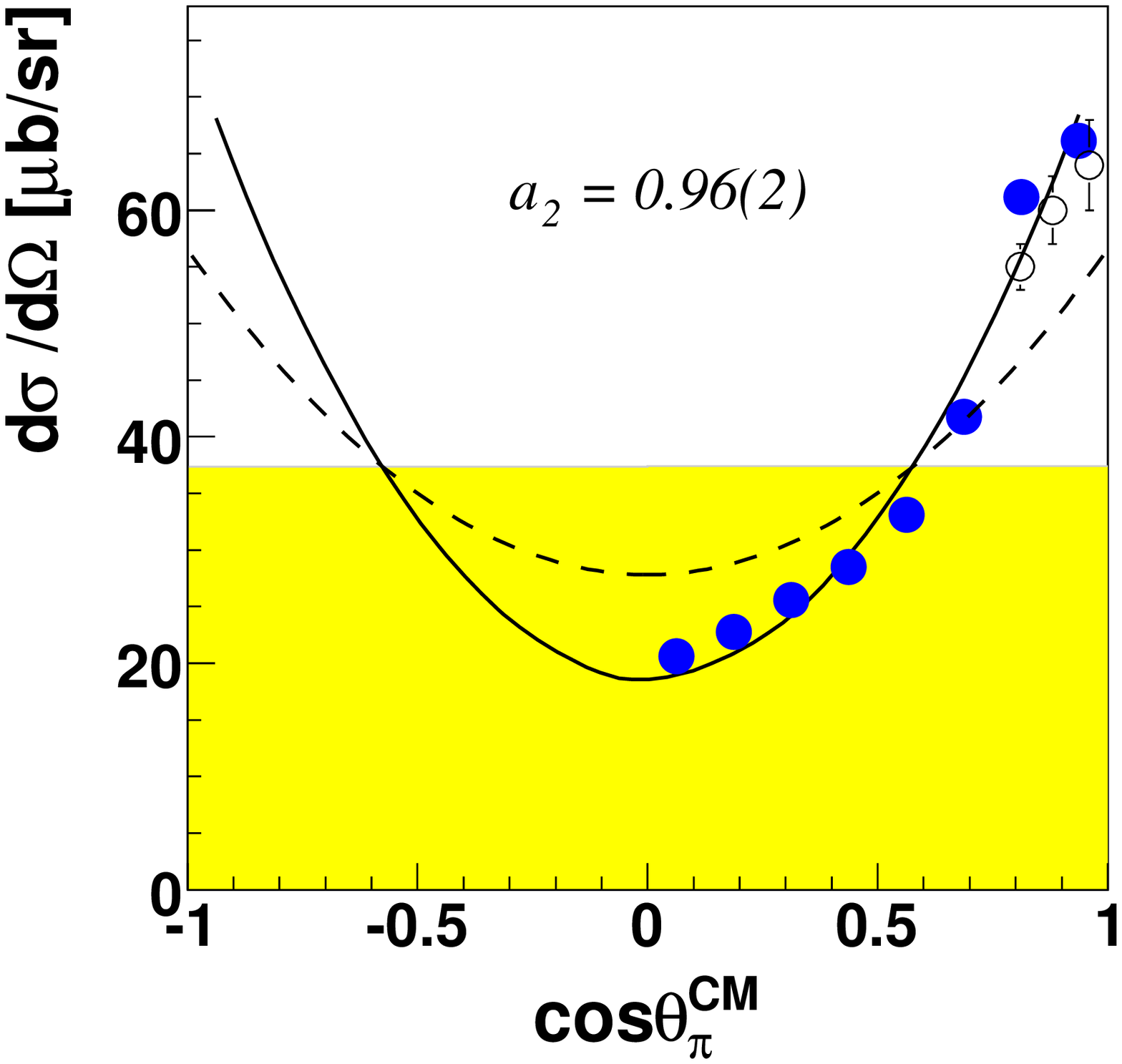}
\end{center}
\caption{Angular distributions of protons (left) and pions (right) in the
  center-of-mass system for the $pp\rightarrow pn\pi^+$
  reaction. Data of this work are shown by full circles, WASA/PROMICE results
  \cite{AB} by open circles (renormalized to $\sigma_{tot}$ of this work) and
  phase space by the
  shaded area. Solid and dashed lines denote s-channel and t-channel
  calculations, respectively, as discussed in the text. For ease of comparison
  the calculations have been normalized to the experimental total cross
  section. Given are also the Legendre coefficients $a_2$ from a Legendre fit
  according to eq. 2 in (I) tgether with statistical errors.
}
\end{figure}

In contrast to the $pp\pi^0$ channel the Dalitz plots for the $pn\pi^+$ channel
are far from being flat -- see Fig.4. This is also borne out in the
projections, the invariant mass spectra $M_{p\pi^+}$ and
$M_{n\pi^+}$ (see Fig. 5, top), which
peak at the highest kinematically available masses. In contrast to these
the  $M_{pn}$ spectrum (Fig. 5, bottom) peaks at the lowest masses. Note that
the high-mass enhancement is similar in both $M_{p\pi^+}$ and $M_{n\pi^+}$
spectra. This is in agreement with the trend observed in bubble chamber data
at $T_p$ = 432 MeV \cite{shim1}. 

Whereas the proton angular distribution is close to flat,
the pion angular distribution is strongly anisotropic. It is fitted very well
by a pure  $(3 cos^2\Theta^{cm}_{\pi^+} +1)$ distribution (solid line in Fig.3,
right).  

Our results are in good
agreement with the results from PROMICE/WASA \cite{AB}, which are the only
other unpolarized data available at $T_p \approx$ 400 MeV. Note, however, that
those measurements were carried out in a very 
limited phase space region only. E.g., the pion
angular distribution was only measured for $cos(\Theta^{cm}_{\pi^+}) \geq$
0.79.

\section{Discussion of Results}

\begin{figure}
\begin{center}
\includegraphics[width=10pc]{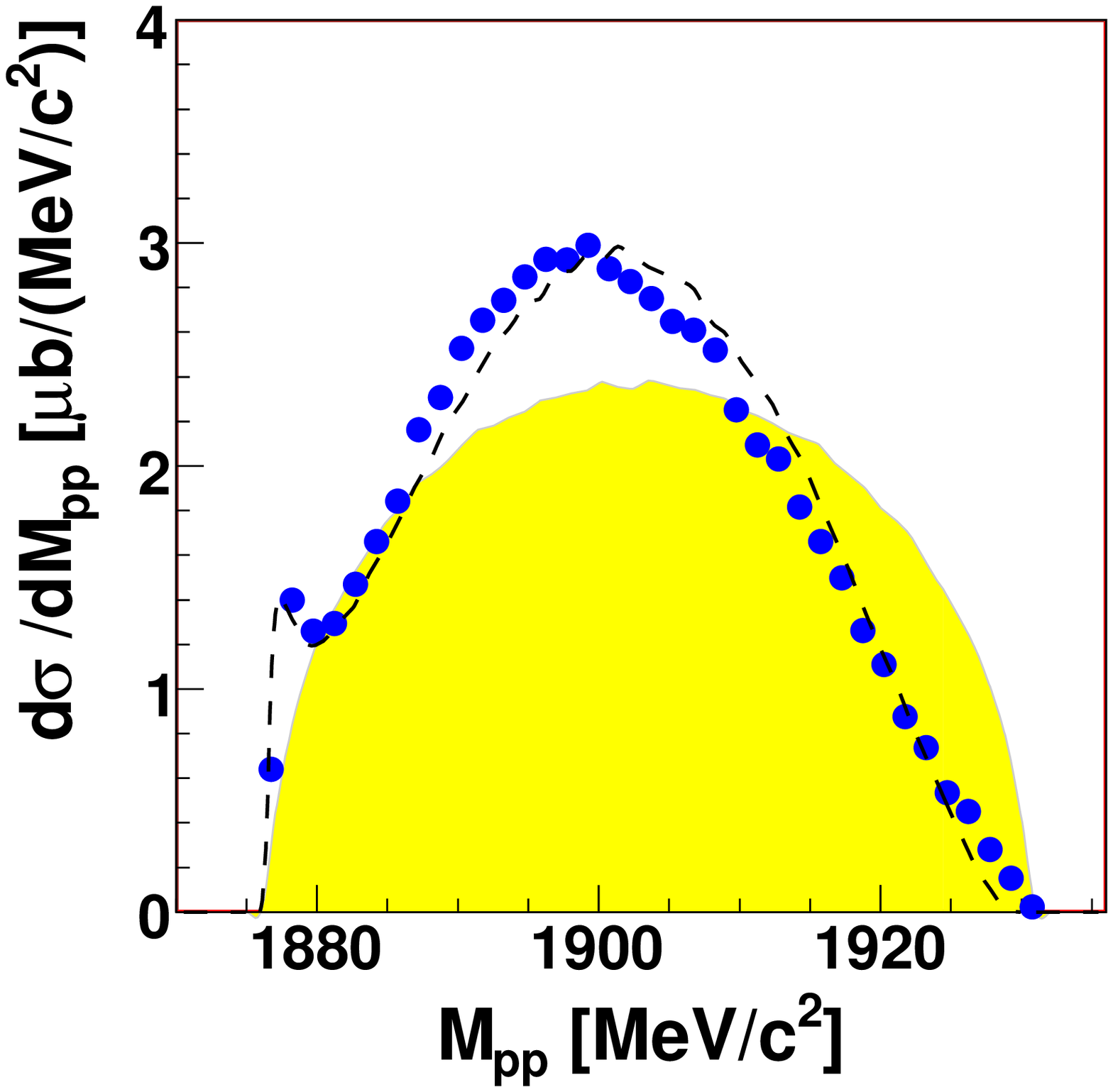}
\includegraphics[width=10pc]{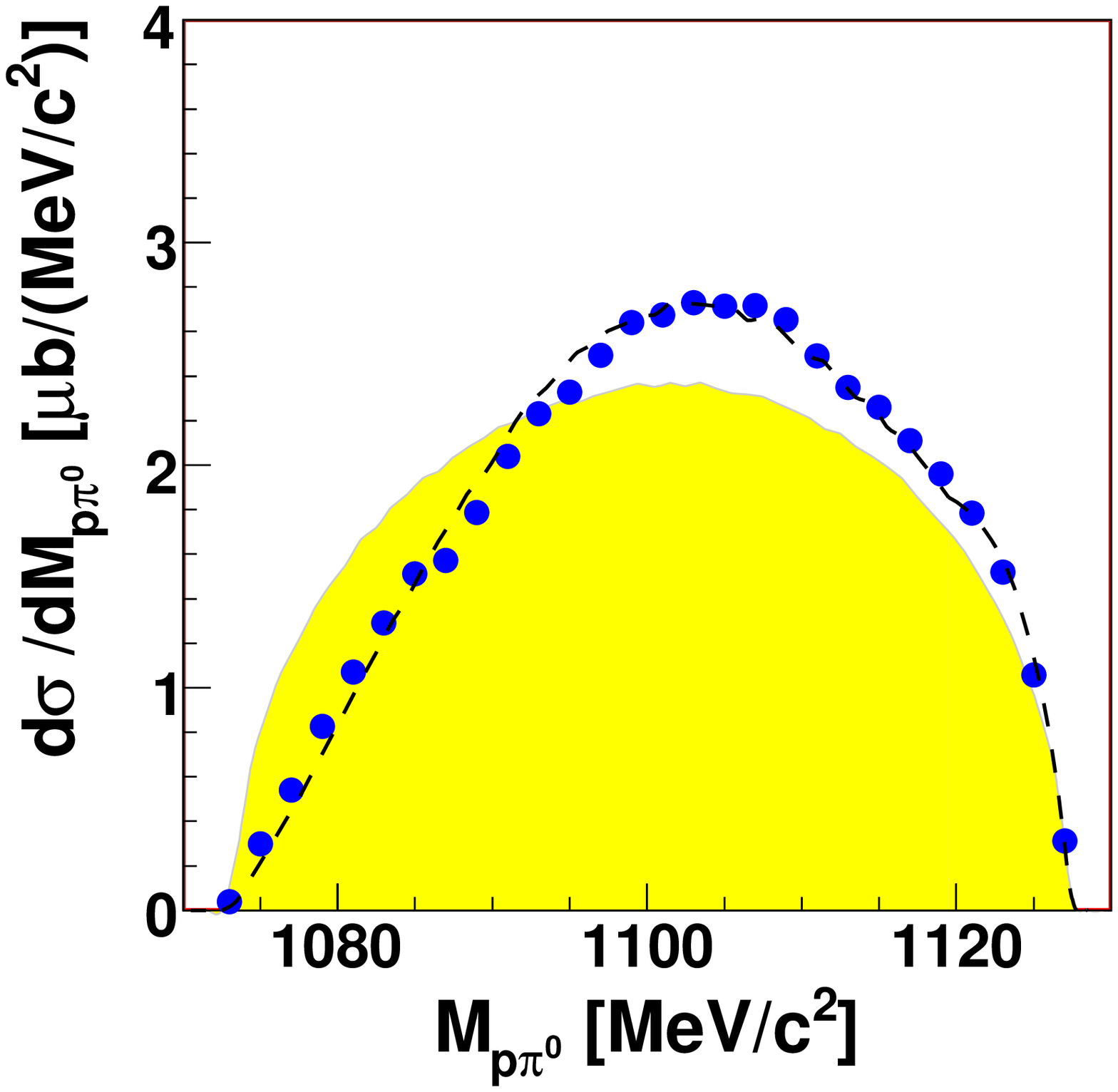}
\end{center}
\caption{Differential cross sections in dependence on invariant masses
  $M_{pp}$ (left) and $M_{p\pi^0}$ (right) for the $pp\rightarrow pp\pi^0$
  reaction (from (I)). Phase space is shown by the shaded
  area. For the explanation of the curves see  (I).} 
\end{figure}

\begin{figure}
\begin{center}
\includegraphics[width=10pc]{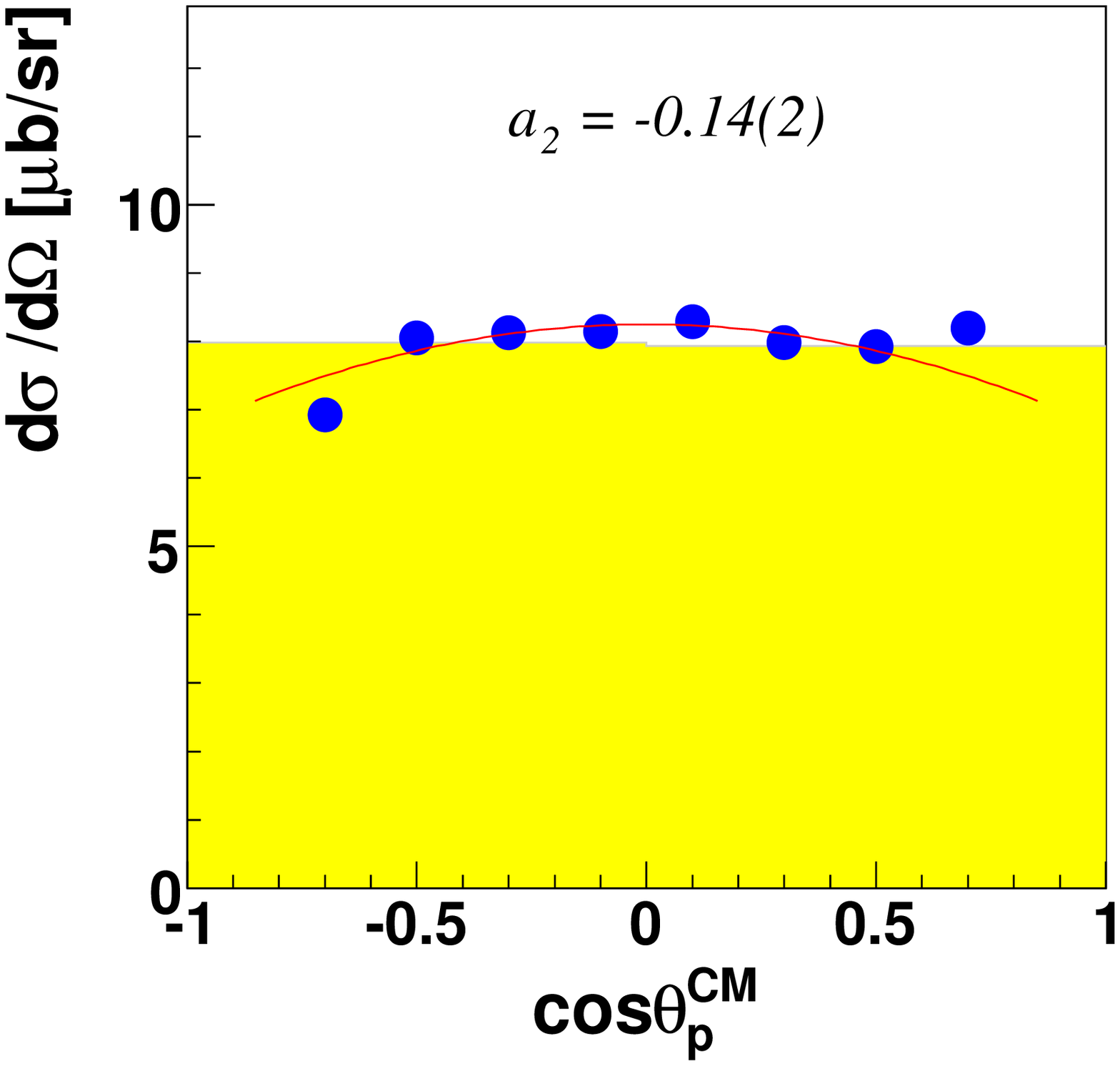}
\includegraphics[width=10pc]{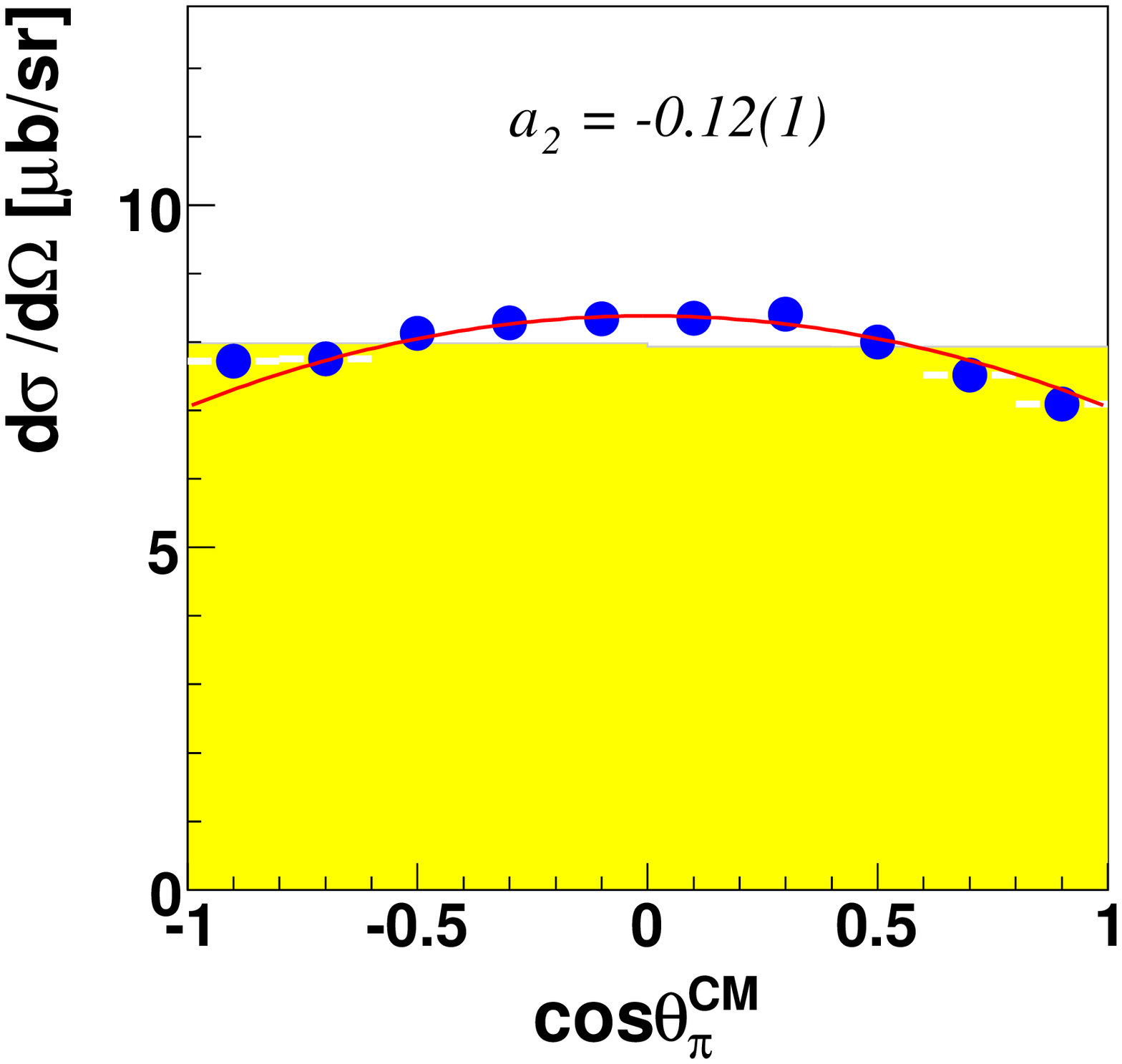}
\end{center}
\caption{Angular distributions of protons (left) and pions (right) 
  in the center-of-mass system for the $pp\rightarrow pp\pi^0$ reaction (from
  (I)). The solid curves represent Legendre fits, see (I). }
\end{figure}

\subsection{Comparison with the $pp\pi^0$ channel}

For the comparison of $pn\pi^+$ and $pp\pi^0$
channels we show once more results in Figs. 7 and 8, which we obtained
previously for the $pp\pi^0$ channel, see (I).

The proton angular distributions both in $pn\pi^+$ and $pp\pi^0$ channels are
close to isotropic (see Figs. 6 and 8, left). They
nevertheless show a tendency of opposite curvature,
which just may be some reflection of the very different pion angular
distributions in both channels. At these low energies the pion angular
distributions are described conveniently by the Legendre ansatz

$\sigma(\Theta_{\pi^0}^{cm}) \sim 1 + a_2 *(3cos^2\theta_{\pi^0}^{cm} - 1) /
2$,

where the parameter $a_2$ is fitted to the data. This ansatz has also been
used in (I) for the analysis of the $pp\pi^0$ channel -- see eq. 2 in
(I). Other notations related to our $a_2$ parameter are as follows: in the
notation of Ref. \cite{pley}, Table III we have $a_2 = A_2 / A_0$, in the
notation of Ref. \cite{har}, Table V we have $a_2 = B / A$ and in the notation
of Ref. \cite{WD}, Table II we have $a_2 = 2 * b_{00}$ (since $a_{00} = 1$
there).       

For the pion angular distribution in case
of $pp\pi^0$ we find a slightly concave shape (Fig. 8, right; $a_2$~=~-0.12),
whereas we get a strongly convex shape in case of the $pn\pi^+$ 
channel (Fig. 6, right; $a_2$ = +0.96(2)).

A striking difference to the $pp\pi^0$ channel is the dominance of the 
$\Delta$  
excitation in the $pn\pi^+$ channel. In the $pp\pi^0$ channel $\Delta$
production in a s-wave relative to the accompanying nucleon is prohibited by
quantum numbers. However, as demonstrated in (I), the $\Delta$ production in
relative p-wave is allowed and indeed contributes significantly to this
channel already close to threshold.

The signature of $\Delta$ production in the $pn\pi^+$ channel is seen in all
differential observables shown in Figs. 3 - 6. The $M_{pn}$ spectrum (Fig. 5,
bottom) is markedly different 
from the $M_{pp}$ spectrum in the $pp\pi^0$ channel (Fig. 7, left). Though the
$pn$ final 
state interaction (FSI) is somewhat larger than the $pp$ FSI, it cannot
account for the huge difference between both spectra at low masses. It
rather is a reflection from the complementary spectra $M_{p\pi^+}$ and
$M_{n\pi^+}$, which peak just at the highest available masses (Fig. 5,
top) in contrast to the more phase-space like $M_{p\pi^0}$ spectrum observed
in the $pp\pi^0$ 
channel (Fig. 7, right). The high-mass enhancement in turn is most likely 
associated with $\Delta$ excitation.  

\subsection{Pion angular distribution}

The most striking feature in our $pn\pi^+$ data at $T_p$ = 400 MeV is the
$(3~cos^2\Theta^{cm}_{\pi^+}  + 1)$ dependence of the pion angular
distribution. It is exactly this
dependence, which is seen in $\pi N$ scattering in the $\Delta$ resonance
region and which is also observed in the $pp \to d\pi^+$ reaction over a large
range of incident 
energies. We note that also the pion angular distributions at $T_p$ = 420 and
500 MeV, deduced from inclusive measurements of the $pp \to pn\pi^+$ reaction
in Ref. \cite{pley} are basically in accord with this dependence.

We also note in passing, that in contrast to the situation in the $pp\pi^0$
channel, where we observe a strong dependence of the pion angular distribution
on the relative momentum q between the two nucleons in the exit channel, we
do not find such a behavior here. The pion angular  distributions stay
essentially the same for different q-regions. This again is in support of a
single dominant mechanism proceeding via a single partial wave.

Whereas we find good agreement in the anisotropy of the pion angular
distribution with the ones measured at TRIUMF at  $T_p$ = 420 and 500 MeV
\cite{pley}, we 
face a profound discrepancy with  IUCF results \cite{WD}. There 
kinematically complete measurements with both polarized beam and target are
presented for $T_p$ = 325, 350, 375 and 400 MeV together with Legendre
polynomial fits, which reproduce the measured 
angular distributions of the polarization observables. In Table II of
Ref. \cite{WD} the coefficients resulting from these fits are listed. This
list also contains the anisotropy parameter $b_{00} = a_2 / 2$ for the
unpolarized differential cross section $\sigma(\Theta_\pi^{CM})$. Since that
paper concentrates on the polarization observables, it does not dicuss the
unpolarized $\sigma(\Theta_\pi^{CM})$ data in any detail. It is just noted
that these have been deduced from $p + n$ coincidences. As a result of their
Legendre analysis they obtain $a_2$ = 0.39 (9) for $T_p$ = 400 MeV, a value,
which is less
than half of our value. The difference amounts to six standard deviations. We
face here perhaps a similar problem as discussed in (I) for the situation
of the experimental data with the reaction $pp \to pp\pi^0$ at the same beam
energy of $T_p$ = 400 MeV. There also a large discrepancy exists between
the asymmetry parameter $a_2$ found in the IUCF analysis and the one we found
at COSY-TOF. Our result has meanwhile been confirmed by CELSIUS/WASA
data \cite{pia}. In both these works it has been shown that previous
results -- like the one obtained at PROMICE/WASA \cite{zlo} -- suffered from
insufficient phase space coverage combined  with inappropriate
extrapolations into the unmeasured regions.

We note the following surprising trend in the results of the IUCF analyses
concerning the $a_2$ parameter. In agreement with previous IUCF work very
close to threshold \cite{har}, where  
a strong increase in the $a_2$ value with energy is observed ($a_2 = $ 0,
0.12 and 0.31 at $T_p = $ 294, 299 and 319 MeV, respectively), a value of
$a_2$ = 0.34(7) at $T_p$ = 325 MeV is reported in Ref. \cite{WD}, Table
II. This increase just reflects the increasing 
importance of the $\Delta$ excitation in the $pp \to np\pi^+$
reaction. However, above $T_p$ = 325 MeV Ref. \cite{WD}(see again Table II)
finds that the $a_2$ parameter no longer increases, but rather saturates at a
value of 0.39. This implies that the $\Delta$ contribution also suddenly
saturates despite the fact the $\Delta$ pole is still far from being
reached at these energies. Also the J\"ulich model calculations \cite{han},
which are used in Ref. \cite{WD} for comparison with the polarization data
predict a 
strongly  increasing importance of the $\Delta$ excitation at these beam
energies -- as also expected intuitively.

In Ref. \cite{WD} the 
discrepancy between their 400 MeV result and the TRIUMF data at $T_p$ = 420 MeV
\cite{pley} was not discussed. The TRIUMF measurements were performed as a
single-arm experiment with a magnetic spectrometer, which measures a
3-body reaction only inclusively. Such measurements are known to be
very reliable for the determination of single-particle angular distributions,
since exactly the same detector is used for the particle detection at all
angles minimizing thus the problem of acceptance and efficiency corrections.

To summarize the discussion about $a_2$ we face a severe discrepancy between
the IUCF and COSY-TOF results at $T_p$ = 400 MeV for both the $pp \to pp\pi^0$
reaction and the $pp \to np\pi^+$ reaction. However, there is good agreement of
our result for the first reaction with the corresponding one from CELSIUS/WASA
and for the second reaction with the one obtained at TRIUMF. We finally note
that the uncertainty of $a_2$ quoted in Fig.6 only includes the statistical
uncertainties of the data. If we include an estimate on systematic
uncertainties (as discussed above in the section 2.4), we
end up with an estimated total uncertainty of $\Delta a_2 \approx$ 0.08.

The close similarity in the angular distributions of $d\pi^+$ and $pn\pi^+$
channels is in agreement with the predictions of F\"aldt and Wilkin
\cite{colin}, according to which the meson angular distributions coincide
closely in bound state and breakup channels near threshold.

The $(3~cos^2\Theta^{cm}_{\pi^+} + 1)$ dependence in the $pp \to d\pi^+$
reaction is due to the $^1D_2$ partial wave in the incident $pp$ channel
combined with the constraint of a p-wave pion relative to the deuteron in the
exit channel. As known from phase shift analyses of this reaction
\cite{said,RA}  
this $^1D_2P$ partial wave dominates the $pp \to d\pi^+$
reaction basically from threshold up to $T_p \approx$ 900 MeV. It is
responsible 
for the resonance-like energy dependence of the total cross section and
performs a perfect looping in the Argand diagram \cite{said}. Thus this partial
wave possesses all features to qualify for a s-channel resonance with $I(J^P) =
1(2^+)$ in the incident $pp$ channel, the final $d\pi^+$ channel and the
intermediate $N\Delta$ channel -- see also discussion in Ref. \cite{hcl}. From
the total cross section of the $^1D_2P$ partial wave in the $pp \to d\pi^+$
reaction \cite{said} we see that the resonance energy is some 60 MeV below the
nominal $N\Delta$ threshold, whereas the resonance width of $\approx$ 110 MeV
corresponds to that of the $\Delta$ resonance.

The experimental observation that up to $T_p$ = 500 MeV essentially a  $(3~
cos^2\Theta^{cm}_{\pi^+} + 1)$ dependence is observed also in the  $pp \to
pn\pi^+$ reaction points to the 
predominance of the $^1D_2P$ partial wave in this reaction, too, as was already
noted in Ref. \cite{pley}. From this also follows that the $pn$ system at these
energies is preferably in the deuteron-like quantum state $I(J^P) = 0(1^+)$.

\subsection{Total  cross section}

The energy dependence of the total cross sections for the $pp \to d\pi^+$ and
$pp \to np\pi^+$ channels is displayed in Fig. 9. For the $pp \to
pn\pi^+$ channel at $T_p$ = 400 MeV there are no experimental data to compare
with. 
However, our result fits well to the trend given by the experimental results
\cite{bys,shim,pley,har} at neighbouring energies. In order to discuss in the
 next section the energy
dependence in some broader context with regard to the dominance of the $^1D_2$
partial wave, we plot in Fig.9 the energy range from threshold up to $T_p$ =
1.5 GeV, i.e. over the full region of $\Delta$ excitation.

\begin{figure}
\begin{center}
\includegraphics[width=22pc]{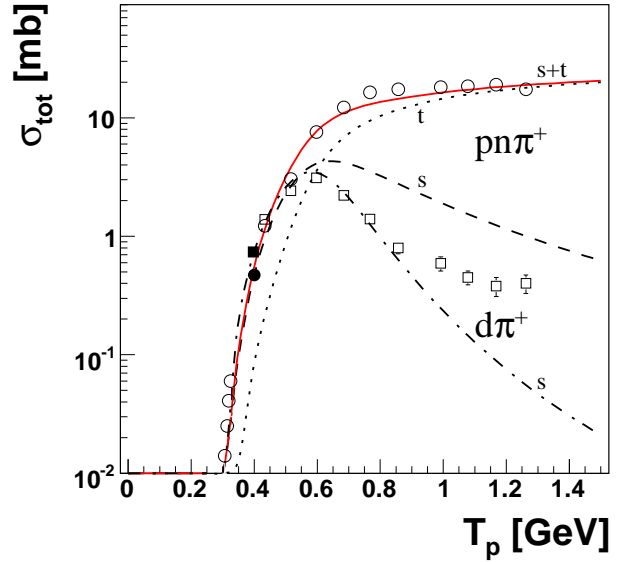}
\end{center}
\caption{Energy dependence of total cross sections for the reactions $pp \to
  d\pi^+$ (squares) and $pp \to pn\pi^+$ (circles). The filled symbols denote
  results of this work, open symbols are from \cite{bys,shim,har}. Dash-dotted
  and dashed lines represent {\it s}-channel calculations for $d\pi^+$ and
  $pn\pi^+$ channels, respectively, normalized to the data at $T_p$ = 400
  MeV. The {\it t}-channel calculation for the $pn\pi^+$ channel is given by
  the dotted line and normalized to the data at $T_p \geq$ 1 GeV, where it is
  known to provide a reasonable description. The
  solid curve denoted by s + t is the incoherent sum of both processes for the
  $pp \to pn\pi^+$ channel.
 }
\end{figure}

\begin{figure}
\begin{center}
\includegraphics[width=22pc]{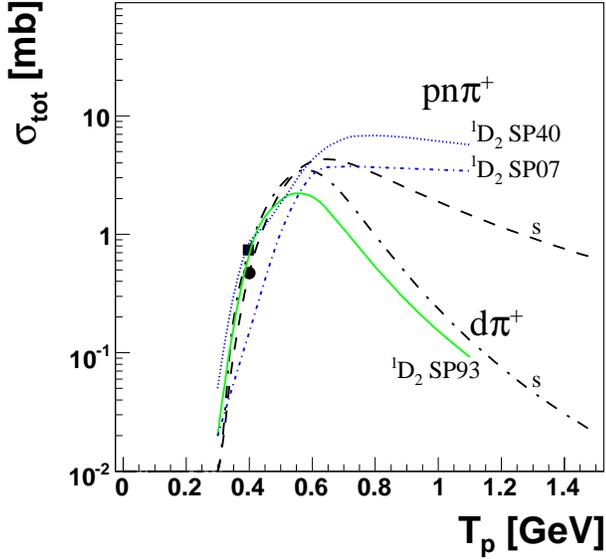}
\end{center}
\caption{Energy dependence of total cross sections as calculated with our
  {\it s}-channel approach for the  $pp \to d\pi^+$ (dash-dotted) and $pp \to
  pn\pi^+$ (dashed cureve) reactions, normalized to the data points from this
  experiment at $T_p$ = 400 MeV (filled square and circle, respectively; see
  also Fig. 9). For comparison the short dash-dotted and dotted lines give
  the SAID  \cite{said} solutions SP07 and SP40, respectively,  for the
  $^1D_2$ partial  wave in the $np\pi^+$ channel (see text) , whereas the
  solid curve shows the
  SAID solution SP93 for the $^1D_2$ partial wave in the $d\pi^+$ channel.
 }
\end{figure}

\subsection{Comparison to {\it t}- and {\it s}-channel calculations}

Finally we confront the data with simple theoretical calculations according to
the graphs depicted in Fig. 1. In this experimental paper it is not our aim to
compare our data with complex calculations, which are beyond the scope of this
work. We rather want to use these calculations in order extract the main
physics message residing in the data. 

Since we have seen in the discussion above that
$\Delta$ excitation and decay is the dominant reaction mechanism and $^1D_2$
is by far the  
dominant partial wave at least in the energy region $T_p$ = 400 - 500 MeV,
we explicitly neglect any non-resonant terms in the reaction process. 

We start with a (properly antisymmetrized) {\it t}-channel approach in
first-order impulse approximation, which
is known to provide reasonable results for the angular distributions at
energies $T_p \geq$ 1 GeV \cite{dmi,aich,fer,mos,sar}. For our energy we show
such a t-channel calculation by the dashed lines in 
Figs. 2, 3, 5 and 6. We see that the pion angular distribution is not
reproduced correctly. The reason for this failure is simple. Since in this
approach the $\Delta$ is excited by pion exchange between the two colliding
nucleons, the reference axis for the pion decay of the excited $\Delta$
resonance is the momentum transfer {\bf q} and not the beam axis. Note that the
latter, 
however, is the reference axis for angles in the center-of-mass frame. Since
the momentum 
transfer varies in dependence of the nucleon scattering angles, which are
integrated over in the pion angular distribution,
the calculated intrinsic $(3~cos^2\Theta^{q}_{\pi^+} + 1)$ dependence in the
{\it t}-channel approach appears to be smeared out in the cms pion angular
distribution.

In order to establish the beam direction as the appropriate reference axis for
the pion angular distribution in the calculation one either has to reiterate
the pion exchanges and sum them up to infinity or simply make a {\it s}-channel
ansatz. The 
latter is easily made by a Breit-Wigner ansatz for the  $^1D_2$ resonance
(Fig. 1, bottom),
which dissociates into a $N\Delta$ system in relative s-wave followed by the
decay of the $\Delta$ resonance. This grants the observed  $(3~
cos^2\Theta^{cm}_{\pi^+} + 1)$ dependence for the pions as well as the
desired shape of the invariant mass spectra. To simulate the  $pp \to d\pi^+$
reaction within this ansatz we impose the Hulthen wave function as condition
for the Fermi momentum between proton and neutron. For the  $pp \to pn\pi^+$
reaction we impose a FSI interaction of Migdal-Watson type \cite{mig,wat}
on the $pn$ system. These calculations (normalized to our results for the total
cross sections at $T_p$ = 400 MeV) are shown by the Dalitz plots in
Fig. 4, bottom and by the solid lines in Figs. 2,3, 5 and 6. They give a good
account of the differential distributions
both for the invariant masses and for the particle emission angles. 

Mass (2.07 GeV) and width (160 MeV) of the $I(J^P) = 1(2^+)$ resonance in this
{\it s}-channel ansatz have been chosen to reproduce the energy dependence of
the total cross section of the  $pp \to d\pi^+$ reaction  in the region of the
$^1D_2P$ dominance, i.e. up to $T_p \approx$ 700 MeV. That way we
simultaneously get a good  
description for the energy dependence of the total cross section for the $pp
\to pn\pi^+$ reaction up to $T_p \approx$ 600 MeV. If combined with the
amplitude of the {\it t}-channel approach we even obtain a reasonable
description of the total cross sections up to the GeV region (Fig.~9).

Finally we compare in Fig. 10 our simple {\it s}-channel calculations for the
$^1D_2$ partial wave in the $np\pi^+$ channel to the results of SAID
\cite{said} partial wave analyses of $NN$ scattering and $pp \to d\pi^+$
reaction. The imaginary part of 
the $^1D_2$ partial wave, which is obtained from the analysis of elastic $NN$
scattering, 
gives rise to the total inelastic cross section $\sigma_{inel}(^1D_2)$ for the
$^1D_2$ partial wave, which in turn is given by the sum of the pion-production
cross sections in this partial wave. Since in the $pp \to pp\pi^0$ reaction
the $^1D_2$ partial wave is highly suppressed due to the suppression of the
$\Delta$ excitation, we have in good approximation 

$\sigma_{pp \to np\pi^+}(^1D_2) \approx \sigma_{inel}(^1D_2) -\sigma_{pp \to
  d\pi^+}(^1D_2)$  

The thus derived results are shown in
Fig. 10 for the SAID solutions SP07 and SP40 from $NN$ scattering analyses
together with the SAID solution 
SP93 for the $^1D_2$ partial wave in the $d\pi^+$ channel. We see that our
{\it s}-channel ansatz overpredicts the $^1D_2$ part of the $pp \to d\pi^+$
cross section 
at higher energies as expected, since we have assumed for simplicity that only
this partial wave contributes. For the $np\pi^+$ channel the SP07 solution
underpredicts the cross section in the near-threshold region strongly, whereas
the SP40 solution, 
which is made for the region $T_p \leq$ 400 MeV, provides a proper energy
dependence, however, is in the absolute scale substantially above the
data and the s-channel calculation in this energy region. At higher energies
up to 800 MeV our calculation is essentially 
between both solutions. This comparison with the SAID solutions for the $^1D_2$
partial wave shows that our simple model description is not in severe
contradiction to the SAID results, where the large spread between both
solutions points to still substantial ambiguities in the $NN$ partial wave
analyses concerning the imaginary parts of partial waves.

\section{Summary}
\label{sec:5}

We have presented measurements of the $pp \to pn\pi^+$ reaction at $T_p
\approx$ 400 MeV. In this energy region they are the first exclusive ones of
solid statistics to cover most of the reaction phase space. The differential
distributions are characterized by the overwhelming dominance of the $\Delta$
excitation. The pion angular distribution coincides with that
observed in the $pp \to d\pi^+$ channel, which in turn is dominated by the
resonating $^1D_2P$ partial wave. From this coincidence we conclude that also
the $pn\pi^+$ channel 
is dominated by the same s-channel resonance in the near-threshold region. The
correlation between boundstate and breakup channels is in accordance with the
predictions by F\"aldt and Wilkin \cite{colin}.

\begin{acknowledgement}

This work has been supported by BMBF, DFG
(Europ. Gra\-duiertenkolleg 683) and COSY-FFE. We acknowledge valuable
discussions with R. A. Arndt, L. Alvarez-Ruso, D. Bugg, C. Hanhart,
M. Kaskulov, V. Kukulin, E. Oset, I. Strakovsky, W. Weise and C. Wilkin.
\end{acknowledgement}

\end{document}